\documentclass[
  letterpaper,
  prx,
  aps,
  twocolumn,
  superscriptaddress,
  nofootinbib,
  nobibnotes
]{revtex4-2}
\usepackage{bm,color}
\usepackage[T1]{fontenc}
\usepackage[utf8]{inputenc}
\usepackage{latexsym}
\usepackage{graphicx} 
\usepackage{amsmath} 
\usepackage{amsfonts} 
\usepackage{braket}
\usepackage{subfigure}
\usepackage{bbold}
\usepackage{lineno}
\usepackage{siunitx}
\usepackage{xcolor}
\usepackage{float}
\usepackage{hyperref}
\usepackage[nameinlink,capitalise]{cleveref}
    \DeclareSIUnit\bar{bar}
    \DeclareSIUnit\gauss{G} 
    \DeclareSIUnit\dBm{dBm}

\hypersetup{
  colorlinks=true,
  linkcolor=blue!45!black,
  citecolor=blue!45!black,
  urlcolor=blue!45!black,
  pdftitle={A phase microscope for quantum gases},
  pdfauthor={J. C. Brüggenjürgen, M. S. Fischer, and C. Weitenberg}
}

\def \be{{\bf e}}

	\newcommand{\bbm}{\begin{pmatrix}}
	\newcommand{\ebm}{\end{pmatrix}}

	\usepackage[normalem]{ulem} 
	\definecolor{mgreen}{RGB}{1,123,0}

\def\be{\begin{equation}}
\def\ee{\end{equation}}

\makeatletter

\def\frontmatter@thefootnote{%
  \ifcase\csname c@\@mpfn\endcsname\relax
    \or\ensuremath{\dagger}%
    \or\ensuremath{\ddagger}%
    \or *%
    \else\@fnsymbol{\csname c@\@mpfn\endcsname}%
  \fi
}

\long\def\frontmatter@footnotetext#1{%
  \par
  \begingroup
    \footnotesize
    \centering
    \setlength{\parindent}{0pt}%
    \@makefnmark\,#1\par
  \endgroup
}

\makeatother

\begin{document}

\title{A phase microscope for quantum gases}

\author{J. C. Brüggenjürgen}
\altaffiliation[Present address: ]{%
  Institut für Experimentalphysik,
  Universität Innsbruck,
  6020 Innsbruck, Austria.
}

\author{M. S. Fischer}
\altaffiliation[Present address: ]{%
  Universal Quantum Deutschland GmbH,
  20457 Hamburg, Germany.
}

\affiliation{%
  Institute for Quantum Physics,
  University of Hamburg,
  22761 Hamburg, Germany
}

\author{C. Weitenberg}

\affiliation{%
  Institute for Quantum Physics,
  University of Hamburg,
  22761 Hamburg, Germany
}

\affiliation{%
  Department of Physics,
  TU Dortmund University,
  44227 Dortmund, Germany
}

\begin{abstract}
Coherence properties are central to quantum systems and are at the heart of phenomena such as superconductivity. Here we study coherence properties of an ultracold Bose gas in a two-dimensional optical lattice across the thermal phase transition. To infer the phase coherence and phase fluctuation profile, we use direct matter-wave imaging of higher Talbot revivals as well as a new phase microscope based on a site-resolved mapping of phase fluctuations to density fluctuations during matter-wave imaging. We observe the algebraic decay of the phase correlations in the superfluid phase and a linear temperature increase of the exponent. These techniques will also allow studying coherence properties in strongly-correlated quantum systems with full spatial resolution.
\end{abstract}

\maketitle

Quantum gas microscopes have emerged as powerful tools capable of resolving density distributions on individual lattice sites, thus enabling precise reconstruction of the density component of the full wavefunction and expanding the notion of experimentally accessible quantities to exotic correlators such as multipoint correlators or string operators \cite{Bakr2010,Sherson2010,Gross2021}. Microscopes are so far mostly limited to the measurement of densities, but cannot access phases locally although proposals for detecting off-diagonal correlations and coherences \cite{Killi2012,Kessler2014,Kosior2014,Ardila2018} and measurements of local currents \cite{Impertro2024} in strongly-correlated systems exist. Measurement of coherence in quantum gases is common practice starting with the interference of condensates \cite{Andrews1997}, extracting coherence from the Bragg peaks of optical lattices \cite{Gerbier2005}, phase fluctuations from short expansion \cite{Dettmer2001} or revival strength in Talbot revivals \cite{Santra2017}, but mostly as global properties probed over the entire system. Local measurements of coherence and phases on the ultimate level of a single lattice site, have so far not been achieved.

Here, we use matter-wave microscopy in a coherent regime to fill this gap and characterize the phase and coherence of a bosonic quantum gas on individual lattice sites in the regime of many atoms per tube. The matter-wave microscope consists of a protocol of evolutions in optical or magnetic traps to achieve a magnified image of the original density distribution \cite{Asteria2021,Zahn2022,Brandstetter2024}. In the lattice experiments, the matter-wave protocol was performed in an incoherent system by freezing out the coherence in a deep lattice before initializing the protocol. Here we consider the matter-wave protocol in a coherent regime and study the resulting Talbot carpet as well as phase-effects on the matter-wave image. We introduce a phase microscope, which gives direct access to phase fluctuation profiles. We combine this with the information from the Talbot carpet, which measures the degree of coherence between the lattice sites \cite{Santra2017}. Finally, we use these techniques to characterize the Berezinskii-Kosterlitz-Thouless (BKT) phase transition in the two-dimensional lattice \cite{Schweikhard2007} via the algebraic decay of the phase coherence. We conclude with an outlook on the various possibilities offered by these new techniques.

\begin{figure}[h!]
    \centering
	\includegraphics[width=0.8\linewidth]{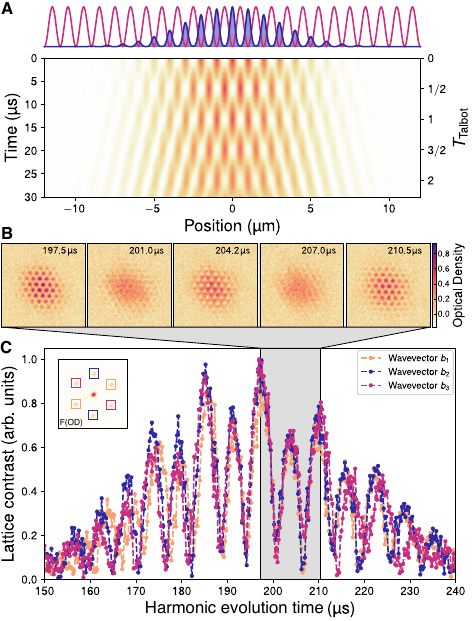}
	\caption{{\bf Matter-wave Talbot effect in a 2D optical lattice.} (A) The Talbot carpet arises from near-field interference of light or matter waves released from a lattice (top). The evolution of the wavefunction in the Gaussian trap shows revivals at multiples of the Talbot time as well as the secondary revivals in between (bottom) (B) Example images of $^7$Li BECs released from a triangular optical lattice of $1.7 E_{\rm rec}$ depth with a matter-wave magnification of 35.3(5) and evolution times of 197.5, 201.0, 204.25, 207.0, and 210.5~$\mu$s, showing the Talbot revivals and the loss of lattice contrast for intermediate times. The images show the optical density (OD) in an area of \SI{8.8}{\mu m} width in the atomic plane. (C) Lattice contrast as a function of evolution time around the imaging condition at \SI{197.5}{\mu s} evaluated along the three directions of the lattice vectors featuring in total 14 Talbot revivals for positive and negative evolution times from the imaging condition (evaluated along the three reciprocal lattice vectors $\boldsymbol{b}_1 $ in orange, $\boldsymbol{b}_2 $ in blue and $\boldsymbol{b}_3 $ in purple). The contrast is evaluated as the signal in the boxes around the Bragg peaks in the Fourier transformed images (see inset).} 
   \label{fig:1}
\end{figure}

{\bf Matter-wave Talbot effect.} 
The Talbot effect describes the repeated self-imaging of a coherently illuminated lattice under evolution in the near field. The resulting Talbot carpet was studied extensively in optics \cite{Wen2013}, but the same effect also applies to matter-wave interference and was studied with cold atoms with nanofabricated material gratings \cite{Clauser1994,Chapman1995,Nowak1997,Abfalterer1997} and optical lattices \cite{Deng1999,Mark2011} as well as molecules \cite{Gerlich2011}. As the sequence of Talbot revivals from an optical lattice depends on the coherence of the initial matter wave over increasing distances, the contrast of the Talbot revivals can be used to study the coherence in the system. This was achieved via Bragg scattering \cite{Miyake2011}, reloading into the optical lattice \cite{Santra2017, Hollmer2019}, and direct imaging in a large spacing 1D lattice \cite{Makhalov2019}.

Here we use the matter-wave optics \cite{Asteria2021} to directly image the different Talbot revivals from a 2D lattice with a significant magnification, which allows to access coherence properties locally (Fig.~1). We work with a Bose-Einstein condensate (BEC) of typically 10,\,000 $^7$Li atoms of mass $m$ in a triangular optical lattice \cite{SupMat}. The three lattice beams have a wavelength of $\lambda_{\rm lat}=\SI{1064}{nm}$, yielding a lattice constant of $a_{\rm lat}=(2/3) \lambda_{\rm lat}$. The matter-wave protocol consists of a lensing pulse of a quarter-period duration in a carefully calibrated harmonic trap of isotropic in-plane frequency $\omega=2\pi \times \SI{1.25}{kHz}$ followed by an approximately free expansion of \SI{5}{ms} resulting in a magnification of 35.3(5) \cite{SupMat}. To avoid interaction effects during the matter-wave protocol and the Talbot evolution \cite{Wei2024,Hollmer2019}, we work with a small scattering length of $3.6 \, a_0$ tuned via a Feshbach resonance. The trap is realized by the lattice beams with a small waist radius of $w_0=\SI{41}{\mu m}$, which limits the system size, but also creates a deviation from a harmonic trap and resulting matter-wave aberrations further discussed below.
The Talbot evolution is realized by scanning the duration of the lensing pulse and is clearly visible as a sequence of revivals of the lattice (Fig.~1C). While the Talbot effect is usually considered in free-space evolution, the same revivals also appear for an evolution in a harmonic trap~\cite{SupMat}. Our data shows the first direct imaging of matter-wave Talbot revivals from a 2D lattice. We evaluate the lattice contrast along the three directions of the triangular lattice and identify 14 revivals in total, counting positive and negative evolution times from the imaging condition (Fig.~1C). The 1st, 2nd and 3rd revivals are progressively probing coherence at longer and longer range \cite{Santra2017}. In our case, the number of revivals is limited by the finite system size with a diameter of about six lattice sites, not by the coherence of the system. 

The Talbot time is given by $T_{\rm Talbot}  = 8\pi^2 m/(h b^2) = 2a_{\rm 1D}^2/(h/m)=\SI{13.2}{\mu s}$, where $h$ is Planck's constant, $b = \sqrt{3}(2\pi/\lambda_{\rm lat})$  the length of the reciprocal lattice vector and $a_{\rm 1D} = 2\pi/b$ the corresponding 1D lattice constant. The observations fit very well to this Talbot time. An evolution by a multiple of $T_{\rm Talbot}$ is called a primary revival and reproduces the original lattice, while an evolution by an odd multiple of $T_{\rm Talbot}/2$ is called a secondary revival and produces a laterally shifted lattice, which in the case of a triangular lattice leads to an inverted lattice similar to a honeycomb lattice \cite{Wen2013}. We indeed see honeycomb-like structures for these secondary revivals (Fig.~\ref{fig:1}C). Such an inverted lattice structure is a distinctive feature of the Talbot effect from higher-dimensional lattices as observed here.

Due to the coherent nature of absorption imaging and the large lattice constant of the magnified density, our system also features an optical Talbot effect visible when scanning the optical imaging plane \cite{SupMat}.

\begin{figure}[t!]
    \centering
	\includegraphics[width=\linewidth]{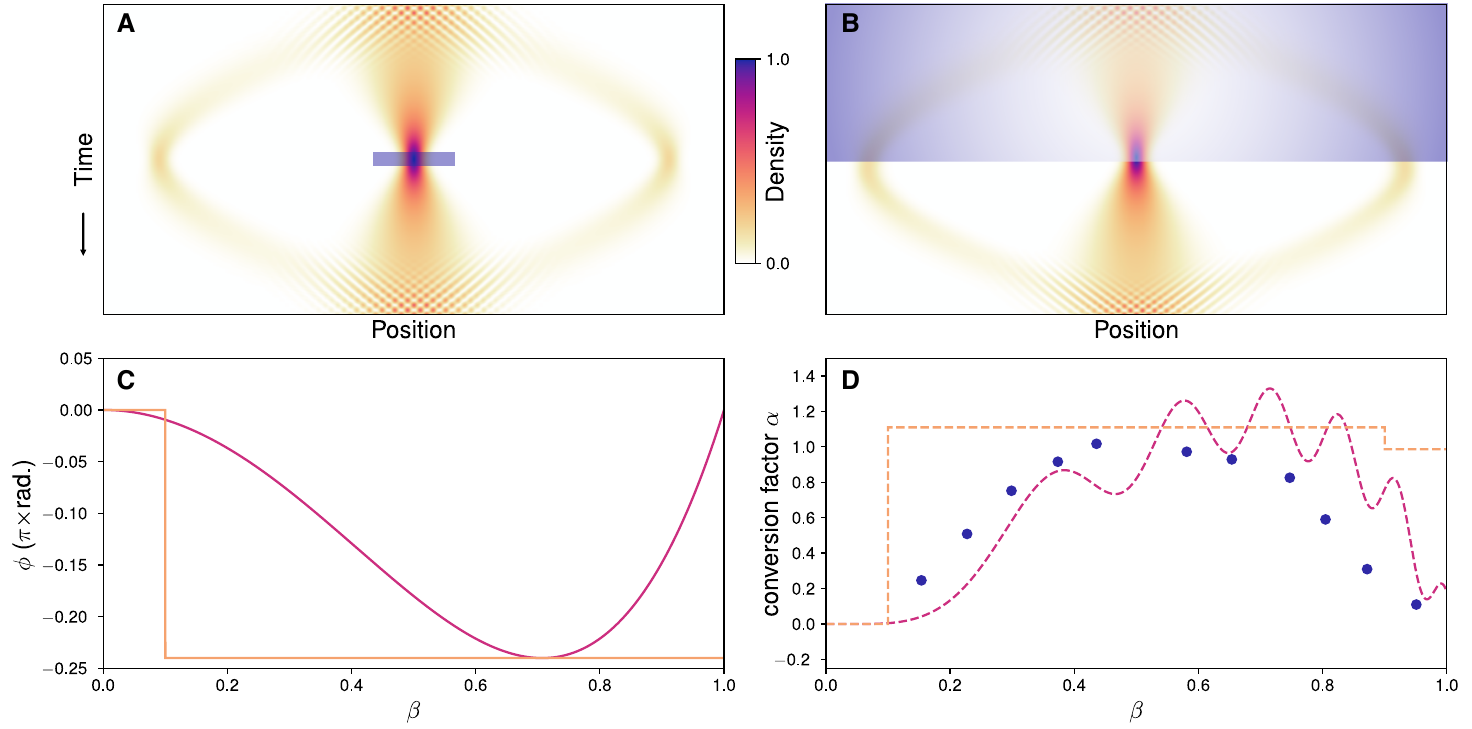}
	\caption{{\bf Aberration-induced phase microscope.} (A) Simulation of the matter-wave density during the matter-wave protocol. A phase microscope imprints a phase shift on the zero momentum component in the Fourier plane of the matter-wave imaging (light blue phase mask), such that the phase fluctuations interfere with it. (B) In the aberration-induced phase microscope, the phase is instead acquired in a continuous fashion due to matter-wave aberrations during the first lensing pulse (blue shading). (C) The acquired phase from the matter-wave aberrations is momentum dependent and goes to zero at the reciprocal lattice vector $b$ (purple line). The phase imprinted by the phase mask is shown for comparison (orange line). (D) The conversion factor $\alpha$ from the initial phase fluctuations to the measured relative density fluctuations as function of momentum is top-hat shaped for the phase mask (dashed orange line, with a small dent at $\beta=1$ due to the last term in Eq.~(2)) and has a similar, but smooth shape for the aberration-induced phase microscope (dashed purple line). The blue circles show results from a numerical simulation for comparison \cite{SupMat}.}
   \label{fig:coherence-microscope}
\end{figure}

{\bf Phase microscope.}
While the Talbot effect can be used to infer coherence properties \cite{Santra2017}, it intrinsically relies on the interference between many lattice sites, making a reconstruction of the initial local phase profile challenging. We therefore realize here a phase microscope inspired by \cite{Murthy2019}, which provides direct access to the local phases by local interference with the zero-momentum BEC, which is selectively phase shifted in the Fourier plane of the matter-wave imaging. 
This can be realized in matter-wave optics by pulsing on a suitable local potential in the center in the Fourier plane realized after the first lensing pulse (Fig.~\ref{fig:coherence-microscope}A). The scheme is analogous to phase contrast imaging in the optical domain \cite{Ketterle1999} and is also relevant for electron microscopes \cite{Axelrod2024} and polariton BECs \cite{Caputo2018}. A similar mapping from phase to density fluctuations also occurs for a short free evolution and was used in continuous quantum gas systems \cite{Dettmer2001,Imambekov2009,Singh2014,Seo2014,Sunami2024,Chen2021}. The key difference here is the use of the matter-wave microscope in imaging condition, where such a mapping occurs without further free evolution and which therefore allows mapping local phase to local density with very high spatial resolution.

The idea of phase contrast imaging relies on approximating the wave function for small phase fluctuations $\theta(x)$ as $\psi(x)\propto e^{i\theta(x)}\approx 1+i\theta(x)$. After imprinting a phase $\phi$ on all momenta relative to the zero momentum component, the wave function becomes $\psi(x)=1+i e^{i \phi}\theta(x)$, giving the second term a real contribution that interferes with the zero-momentum component, when evaluating the density $n(x)=|\psi(x)|^2 \propto 1-2\sin(\phi)\theta(x)$. The conversion factor $\alpha$ between the initial phase fluctuations $\theta(x)$ and the measured relative density fluctuations is then $\alpha=-2\sin(\phi)$.

We consider here a modulated wavefunction in an optical lattice with reciprocal lattice vector $b$, which we model as $\psi(x) \propto e^{i\theta(x)}[c_0+2c_1\cos(bx)]$ with the first two Bloch coefficients $c_0$ and $c_1=c_{-1}$. One can show that the conversion factor becomes $\alpha=-2\sin(\phi)c_0^2$ when considering the densities integrated over the Wigner-Seitz cells and making suitable approximations \cite{SupMat}. We conclude that in a lattice, the mapping still holds, but the conversion factor is reduced because $c_0<1$ and the imprinted phase should be small $\phi \ll \pi/2$ to avoid additional deformations of the density.

Here we realize a variant of this scheme (an aberration-induced phase microscope) which makes use of the naturally present anharmonicity of the optical potential for the lensing pulse and does not require an additional manipulation in the Fourier plane. The aberrations produce a phase shift in a continuous rather than a pulsed fashion (Fig.~\ref{fig:coherence-microscope}B) and the momentum-dependent phase shift profile is smooth rather than top-hat-shaped around zero momentum (Fig.~\ref{fig:coherence-microscope}C), but it produces a conversion factor $\alpha$ with a very similar shape (Fig.~\ref{fig:coherence-microscope}D) as we explain in the following. We note that these aberrations are also present in the measurement of the Talbot revivals discussed above, but they are not important for the qualitative analysis of the contrast. While the phase microscope could be implemented by explicit phase imprinting, for our system with aberrations it occurs naturally when choosing the lensing pulse duration to yield the direct image. We note that Fig.~\ref{fig:1} was taken at low lattice depth of $1.7  \, E_{\rm rec}$, where we have basically no phase fluctuations that would modify the images and Talbot revivals.

We consider the aberrations for a Gaussian optical trap with beam waist radius $w_0$ and note that the aberrations are automatically aligned with the harmonic trap. The aberration-induced phase $\phi_\beta$ for a fluctuation with momentum $\beta b$ with a dimensionless factor $\beta$ is given by
\begin{equation} \label{eq:phi}
    \phi_\beta=\frac{3\pi}{32\hbar}\frac{b^4}{(m\omega_0)^3 w_0^2}(-\beta^2+\beta^4).
\end{equation}
The matter-wave duration is adapted for a sharp imaging of the lattice itself, which is reflected by the condition $\phi_1=0$ \cite{SupMat} (Fig.~\ref{fig:coherence-microscope}C). 

The conversion factor for the momentum-dependent phases is
\begin{equation}
    \alpha_\beta =-2c_0^2 \sin(\phi_\beta)\\-2c_1^2\sin(\phi_{1+\beta}-\phi_1)-2c_1^2\sin(\phi_{1-\beta}-\phi_1).
\end{equation}
For the momentum-dependent phase shift of Eq.~(\ref{eq:phi}), the conversion factor obtains the shape shown in Fig.~\ref{fig:coherence-microscope}D with large values in the range $\beta \in [0.3, 0.9]$. The shape is qualitatively reproduced by numerical simulations \cite{SupMat}. The shape corresponds to a high-pass Fourier filter in contrast to interference techniques with less spatial resolution, which cut away the high spatial frequency fluctuations \cite{Corman2014}.

The situation is qualitatively similar in the experiment, but the 2D lattice makes the exact calculation more involved. 
We calibrate the conversion factor for the experimental situation by a comparison to the expected thermal on-site phase fluctuations at low temperature $T$ given by the variance $\sigma_{\theta}^2(x)\sim k_{\rm B} T/J$ \cite{Tobochnik1979} and obtain $\alpha=0.58(6)$ \cite{SupMat}, where $J$ is the Josephson coupling energy and $k_{\rm B}$ is the Boltzmann constant. It can be assumed constant for our range of lattice depths and for the wave vectors of the phase fluctuations corresponding to distances up to three sites that we evaluate below. This assumption is motivated by this analytic and numerical modelling and justified a posteriori by the comparison of the extracted phase profiles with the BKT physics discussed below. An estimate from the aberrations with $\phi_{\rm max}=-0.24\pi$ and the typical Bloch coefficient $c_0=0.56$ in our experiments yields a conversion factor of $\alpha\approx-2\sin(\phi_{\rm max})c_0^2=0.43$, which matches well with our calibration. 

Phase contrast imaging can also be understood in real space: the phase imprinting on a finite range around zero-momentum endows an initially localized wave function with a long tail with an extent, that is inversely proportional to this finite range. The local phase-to-density mapping then arises from the interference of the respective local wave function with the tails of all other wave functions. For our specific aberration-induced realization of the phase imprinting, which works for a restricted range of wave vectors, this tail is relatively short-ranged and consists of a series of lobes that are spaced by about $0.7 a_{\rm lat}$. Our 1D simulations of the full aberrations yield an intensity of 6\% in the first side lobe. 

{\bf BKT phase transition.} 

\begin{figure*}[ht]
    \centering
    \includegraphics[width=0.95\linewidth]{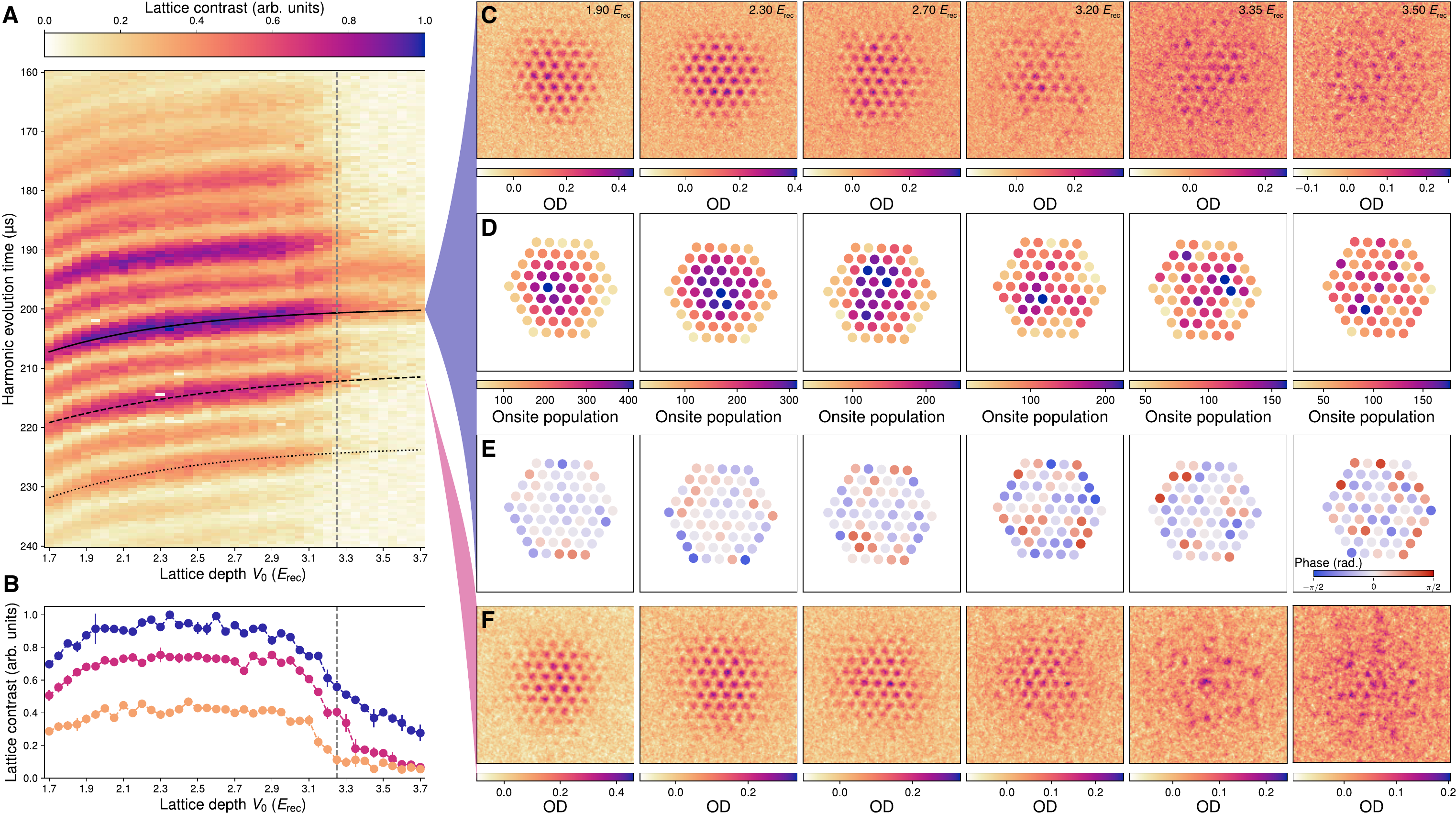}
	\caption{{\bf BKT phase transition in the 2D lattice.} (A) Contrast along the Talbot carpet as a function of the lattice depth. The contrast of the revivals (averaged over the three directions) diminishes for deep lattices. At \SI{6.5}{\mu s} before the first image, close to the negative first secondary revival, a contrast is visible also for deep lattices due to an artifact from the aberrations \cite{SupMat}. At shallow lattices the revivals occur at later times. The left column corresponds to the data shown in Fig.~1C. (B) The contrast along the primary Talbot revivals drops around a critical lattice depth of $V_0=3.25 \, E_{\rm rec}$ (image, first and second primary revival in blue, purple and orange corresponding to the solid, dashed and dotted line in A, respectively). (C) Phase microscope images for lattice depths $1.90, 2.30, 2.70, 3.20, 3.35$ and $ 3.50 \, E_{\rm rec}$ displaying increasing fluctuations for deeper lattices. (D) Measured onsite populations containing the phase fluctuations obtained by integrating the density over the Wigner-Seitz cells. (E) Individual phase profiles obtained by subtracting the mean envelopes and scaling by the conversion factor. The phase fluctuations increase with lattice depth. We estimate a statistical error of the phases from the photon shot noise of the images yielding a standard deviation of 0.08 rad for the central tubes and 0.34 rad for the outermost tubes. (F) Images at the first primary Talbot revival show the loss of lattice contrast for deep lattices and the appearance of speckle images instead.}
   \label{fig:talbot-versus-lattice-depth}
\end{figure*}

The system of a two-dimensional array of tubular Bose-Einstein condensates realized in our optical lattice can be mapped to a classical XY model \cite{Trombettoni2005,Schweikhard2007,Yamashita2019} and is therefore equivalent to a two-dimensional Bose gas, which was used to study the Berezinskii-Kosterlitz-Thouless (BKT) phase transition associated with the unbinding of vortex-anti-vortex pairs to free vortices, which destroy the phase coherence and the superfluidity \cite{Hadzibabic2006,Hung2011,Desbuquois2012,Seo2014,Sunami2024,Murthy2015,Sunami2022}.

We measure the matter-wave Talbot carpet at varying lattice depths and observe that the contrast of the primary revivals drops at a critical lattice depth of $V_{c}=3.25 \, E_{\rm rec}$, associated with the loss of phase coherence (Fig.~\ref{fig:talbot-versus-lattice-depth}A,B). The positions of the revivals shift to $3\%$ longer evolution times for shallower lattices, which we attribute to interaction effects during the matter-wave protocol for our small but finite scattering length. Fig.~\ref{fig:talbot-versus-lattice-depth}F shows exemplary pictures of the first primary revival for increasing lattice depths. For shallow lattices the pictures show clear revivals of the original lattice, while at deeper lattices, we observe patterns reminiscent of the laser speckle as they form under diffraction of coherent laser light from random media. This is the expected behavior for near-field interference from quasi-random phases at each tube, but well defined individual BECs in each tube \cite{Makhalov2019}.

Exemplary pictures of the original matter wave image for increasing lattice depths are plotted in Fig.~\ref{fig:talbot-versus-lattice-depth}C. We find that the system diameter increases from 5 to 9 lattice sites, which we attribute to the increasing interaction strength from the tighter confinement on the lattice sites. We fit the mean onsite density (Fig.~\ref{fig:talbot-versus-lattice-depth}D) of 100 such images with a Gaussian envelope and plot the relative density fluctuations averaged over the Wigner-Seitz cells in Fig.~\ref{fig:talbot-versus-lattice-depth}E. The XY model is a great use case of the phase microscope introduced above, because the initial density fluctuations are suppressed \cite{Hadzibabic2011} and the measured density fluctuations can clearly be identified with initial phase fluctuations. The pictures show increasing phase fluctuations and different phase domains for deeper lattices. This direct imaging of the phase fluctuations with single-site resolution is a central result of our work. The characterization of the system relies on the combination of the Talbot revivals and the phase microscope. The Talbot measurements establish the phase coherence between the tubes for shallow lattices, i.e., the presence of perturbative phase fluctuations on a coherent background, which we use for the interpretation of the phase microscope images.

The control parameter for the phase transition is the ratio $k_{\rm B}T/J$, which can be controlled via the lattice depth \cite{Schweikhard2007}. We obtain the Josephson coupling energy $J=2KN_i$ \cite{Trombettoni2005} from a band structure calculation for the single-particle tunneling energy $K$ and from the microscope images for the atom number per tube $N_i$. A temperature determination from a bimodal fit to the images as in ref.~\cite{Asteria2021} is not feasible in our very low-temperature regime. We therefore extract the temperature from the phase thermometry also used above for the conversion factor \cite{SupMat}. This calibration is a lattice generalization of the sensitive phase thermometry previously introduced for a BEC in a double well potential \cite{Gati2006}. The temperature curve is fixed by the BKT phase transition at $V_{c}=3.25 \, E_{\rm rec}$, which corresponds to a ratio of $k_{\rm B}T/J=1.43$ \cite{Okabe2025} and the extracted temperatures lie in the range between 19.0(5) and 61(2)\,nK \cite{SupMat}.

\begin{figure}[t]
    \centering
	\includegraphics[width=0.9\linewidth]{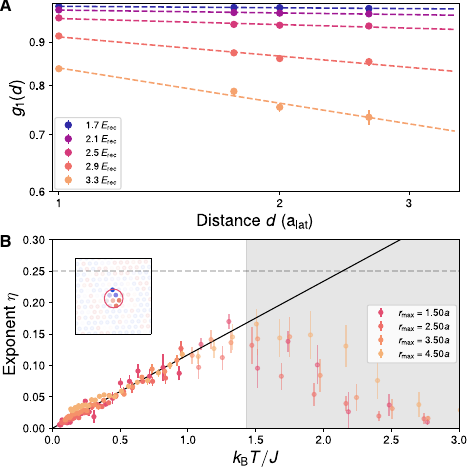}
	\caption{{\bf Quasi-long-range order in the superfluid regime.} (A) The phase coherence function $g_1(d)$ as a function of the distance $d$ evaluated from the phase microscope shows an algebraic decay (dashed lines) and we fit an exponent $\eta$. The shown values are extracted for lattice depths $V_0 = 1.7, 2.1, 2.5, 2.9$ and $3.3\,E_{\rm rec}$ (from top to bottom) for a disk radius of $1.5\,a_{\rm lat}$. The phase coherence function $g_1(d)$ is obtained by averaging the phase differences between pairs of lattice sites with distance $d$ inside the disk. (B) The exponents $\eta$ evaluated using all lattice sites within disks of increasing radius ($1.5,2.5,3.5$ and $4.5 \,a_{\rm lat}$ (dark to light orange)) collapse onto a single line, when normalizing the temperature by the respective tunnel element using the atom number averaged over the disk. The data follows the expected linear increase with $k_{\rm B}T/J$ in the superfluid regime (non-shaded area) and we fit a slope using data up to $k_{\rm B}T/J=1$ (black line), which yields $\eta_c=0.17(3)$ at the phase transition to the normal phase (shaded area). The inset illustrates the disk of radius of 1.5 sites used in (A).}
   \label{fig:g1-function}
\end{figure}

{\bf Evaluation of quasi-long-range order.} 
For the evaluation of the phase profiles of the phase microscope, we determine the phase coherence function $g_1(d)=\left\langle \psi^\dagger(\boldsymbol{r}) \psi({\boldsymbol{r} + d}) \right\rangle \propto \exp(-1/2 \left\langle(\Delta\theta)^2({d})\right\rangle)$ \cite{Hadzibabic2011} from the phase differences $\Delta\theta=\theta(\boldsymbol{r})-\theta(\boldsymbol{r} + d)$ at distances of $d$ lattice sites, evaluated by averaging over positions within different radii and over 100 images. Here $\psi({\boldsymbol{r}})$ is the wavefunction at position ${\boldsymbol{r}}$. Typical results are plotted in Fig.~\ref{fig:g1-function}A in a log-log plot and show the expected algebraic decay $g_1(d)\propto d^{-\eta}$ in the superfluid BKT phase. The identification as algebraic decay is further supported by the comparison to an exponential fit, which consistently has a $\chi^2$ value a factor three larger.

We plot the resulting exponents $\eta$ in Fig.~\ref{fig:g1-function}B as a function of $k_{\rm B}T/J$ using the average atom number within the radius for the determination of $J$. The collapse of the curves in the superfluid regime supports the validity of the algebraic decay in trapped systems. For larger temperatures, the stronger fluctuations are not properly mapped by the phase microscope leading to a shallow $g_1(d)$ profile and small exponents under the invalid assumption of algebraic decay. In the superfluid regime, the exponents follow the Nelson Kosterlitz prediction of a linear scaling with temperature \cite{Nelson1977,Sunami2022} and we extract the critical exponent at the phase transition of $\eta_c=0.17(3)$, where the error stems from the systematic error on the conversion factor. 

This exponent at the phase transition of $\eta_c=0.17(3)$ is close to the universal value of $\eta_c=0.25$ of the infinite homogeneous case. An analysis of harmonically trapped systems based on spin wave approximation \cite{Boettcher2016} also found that the coherence function still decays algebraically, but that the exponent is increased relative to the infinite homogeneous case, while it is predicted to decrease for finite homogeneous systems \cite{Weber1988}. Previous experiments with ultracold atoms in harmonically trapped continuous systems measured a value of $\eta_c=0.17(3)$ for a real-space evaluation with finite slice width \cite{Sunami2022} and $\eta_c=1.4$ for a global momentum-space evaluation of a system across the BEC-BCS crossover \cite{Murthy2015}. This illustrates that in finite trapped systems the physics can depend on details of the system and further theory work as well as experiments with local probing of the phase fluctuations such as with the phase microscope introduced here are required.

Our analysis shows that the phenomenology of the BKT phase including the algebraic decay and the linear temperature dependence of the exponent can be identified already in small trapped systems. At the same time, the analysis of the phase fluctuations serves as a validation of the phase microscope. In the future, the system size could be increased and the system will then act as a quantum field simulator, which could be used to explore beyond mean-field physics in the XY model. One could also image the phases in an antidot lattice, where reaching a regime of dominant quantum fluctuations might be more feasible \cite{Yamashita2019}.

{\bf Conclusions and Outlook.} In conclusion, we have realized a phase microscope for quantum gases based on matter-wave optics and used it to probe the coherence properties and phase fluctuations of a Bose gas in a two-dimensional lattice across the BKT phase transition. The possibility to image phase fluctuation profiles and coherence properties with spatial resolution is crucial in inhomogeneous systems such as trapped systems or quasicrystals \cite{Viebahn2019}, but also for detecting chiral domains \cite{Ozawa2023}. Experiments in larger systems will allow probing scaling exponents in Kibble Zurek experiments \cite{Navon2015}.

The matter-wave microscope allows imaging well below the lattice constant \cite{Asteria2021} pointing to in-situ detection of lattice orbitals. The phase microscope introduced here could also be pushed to detect phase structures within the lattice sites \cite{Ikonnikov2020}, such as phase vortices that can arise within higher orbitals due to spontaneous symmetry breaking \cite{Wang2021}. 

Finally, we expect that the techniques demonstrated here can be extended to the regime of single-atom resolved imaging as a full-fledged phase microscope. Implementing a fast switching of the interactions for the matterwave protocol \cite{Holten2022,Brandstetter2024} will allow to probe strongly-correlated systems with the techniques demonstrated here. Recent experimental work used the time evolution in double wells of a superlattice to measure current operators in the strongly correlated regime~\cite{Impertro2024}. The more general manipulation in Fourier space pioneered here might enable even more general measurements involving long-range correlations. A full-fledged phase microscope would allow extracting correlations between local phases and coherences bringing the characterization of quantum many-body systems to a new level. It could be employed, e.g., for distinguishing superfluid phases from Bose glass phases \cite{Soyler2011}, for local probing of coherence in trimerized Mott insulators \cite{Barter2020}, or for identifying ergodic bubbles in many-body localized systems \cite{DeRoeck2017}. 

In our study, the Talbot revivals were important to establish the coherence of the system, which is assumed in the analysis of the phase microscope. In the future, it would be interesting to realize a Talbot microscope via phase retrieval algorithms applied to images at a Talbot revival. The interference of many neighbouring sites at the Talbot revivals makes this analysis more involved, but it potentially allows to measure large phase fluctuations, the degree of coherence, and work in the strongly-correlated regime. As phase retrieval is an inverse problem, we envision that machine learning methods \cite{Carleo2019} can help to realize the Talbot microscope.

\begingroup
\raggedright
\noindent
Corresponding author: Christof Weitenberg.\\
Email:
\href{mailto:christof.weitenberg@tu-dortmund.de}
{\nolinkurl{christof.weitenberg@tu-dortmund.de}}
\par
\endgroup

{\bf Acknowledgements} We thank Klaus Sengstock and Luca Asteria for fruitful discussions and Andreas Kerkmann, Michael Hagemann, Benno Rem and Niklas Käming for contributions in an early stage of the project. We thank Jan Kierfeld for kindly providing the calculation of the thermal fluctuations.
\paragraph*{Funding:}
The work is funded by the European Research Council (ERC) under the European Union’s Horizon 2020 research and innovation program under Grant Agreement No. 802701 and by the Cluster of Excellence “CUI: Advanced Imaging of Matter” of the Deutsche Forschungsgemeinschaft (DFG)—EXC 2056 —Project ID No. 390715994.
\paragraph*{Author contributions:}
J.C.B. and M.S.F. measured and analyzed the data. J.C.B performed numerical simulations. C.W. supervised the work. All authors contributed to planning the experiment, discussions, theoretical analysis, and preparation of the manuscript. 
\paragraph*{Competing interests:}
The authors declare no competing financial interests.
\paragraph*{Data and materials availability:}
Experimental data and analysis code are available at \cite{TUDOdata}.


%


\end{document}


\title{Supplementary Material: A phase microscope for quantum gases}

\author{J. C. Brüggenjürgen}
\altaffiliation[Present address: ]{%
  Institut für Experimentalphysik,
  Universität Innsbruck,
  6020 Innsbruck, Austria.
}

\author{M. S. Fischer}
\altaffiliation[Present address: ]{%
  Universal Quantum Deutschland GmbH,
  20457 Hamburg, Germany.
}

\affiliation{%
  Institute for Quantum Physics,
  University of Hamburg,
  22761 Hamburg, Germany
}

\author{C. Weitenberg}

\affiliation{%
  Institute for Quantum Physics,
  University of Hamburg,
  22761 Hamburg, Germany
}

\affiliation{%
  Department of Physics,
  TU Dortmund University,
  44227 Dortmund, Germany
}

\maketitle

\section{Optical Talbot effect}\label{sec:optical-talbot}

Due to the coherent nature of absorption imaging, one also has to consider an optical Talbot effect which arises when moving the plane to be imaged. For the lattice constant magnified by $M_{\rm mw}=35.3(5)$, the optical Talbot length becomes $L_{\rm Talbot} = 2(M_{\rm mw} a_{\rm 1D})^2/\lambda_{\rm im} = 1.4$\,mm, where $\lambda_{\rm im}=671$\,nm is the wavelength used for the absorption imaging of the atoms. With the Talbot length in the millimeter range, the Talbot revivals are not washed out by the transverse extension of the cloud itself or the depth of focus of the imaging system as in the case of direct in-situ imaging, where no Talbot effect was observed upon scanning the image plane \cite{Weitenberg2011PRL}. The optical Talbot effect was also observed from the large interference fringes in an atom interferometer and used to enhance their apparent visibility \cite{Zhai2018}. 

\begin{figure}[ht]
    \centering
	\includegraphics[width=0.9\linewidth]{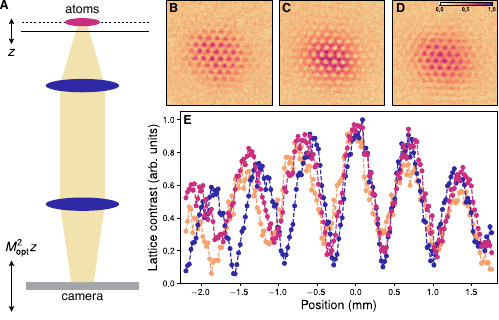}
	\caption{{\bf Optical Talbot effect.} (A) By scanning the camera position, different planes are imaged. (B-D) Images of the BEC in the lattice for the fixed matter-wave imaging condition, but varying positions of the optical imaging plane yielding primary and secondary optical Talbot revivals. (B) At $z=0$ the image of the magnified lattice is captured. (C) The secondary revival at $z \approx \SI{0.7}{mm}$ and (D) the primary revival at $\approx \SI{1.4}{mm}$ fit well with the expected value of the optical Talbot length $L_{\rm Talbot} = \SI{1.4}{mm}$. (E) Lattice contrast as a function of the image plane showing six revivals (evaluated along the reciprocal lattice vectors $\boldsymbol{b}_1 $ in orange, $\boldsymbol{b}_2 $ in blue and $\boldsymbol{b}_3 $ in purple). The slight dephasing between the revivals along the different directions is due to a small deviation of the lattice angles.}
   \label{fig:optical-talbot}
\end{figure}

We image at different planes $z$ by scanning the camera position by $M_{\rm opt}^2 z$ around the image plane, where $M_{\rm opt}=3$ is the optical magnification of the imaging system (Fig.\,\ref{fig:optical-talbot}A). To determine the contrast, we integrate the Fourier signal over the Bragg peak and then take the absolute value in order to avoid finite-size effects that otherwise produce a finite contrast for all camera positions. For our parameters with initial wavepackets that are not localized much stronger than a lattice constant, the Talbot carpet consists only of the primary and secondary revivals (compare Fig.~1A). Their Bragg peak signals have slightly different positions and are out of phase in the overlapping region between the revivals, such that they cancel when integrating before taking the absolute value, yielding zero contrast between the revivals. Monitoring the contrast of the lattice along the three reciprocal lattice vectors, we clearly observe several primary and secondary Talbot revivals (Fig.~\ref{fig:optical-talbot}E). The measured optical Talbot lengths are in excellent agreement with the expectation. The original image (zeroth revival) is clearly identified as the position where the contrast maxima in all three directions are strongest and fall on top of each other. We work at this camera position for all other measurements described in the manuscript.

In the optical Talbot revivals, the maxima along the three directions are not reached for exactly the same position. This is due to a deviation from the 120$^\circ$ arrangement of the laser beams by about 2$^\circ$, which is also visible in the imaged lattice structure itself. The Talbot lengths are therefore slightly different along the three directions. The same effect also applies to the matter-wave Talbot effect, where the revivals along the three directions also dephase for higher Talbot revivals (compare Fig.~1).

Optical imaging is coherent, but the coherence comes from the laser light and the imaging ignores the phase of the atomic wave function. Therefore phase fluctuations of the atomic wavefunction do not affect the optical Talbot revivals as in the phase microscope. Nevertheless, one could in principle confuse a primary revival of either the matter-wave or optical Talbot effect for a combination of secondary revivals in both Talbot effects. Secondary revivals change the lattice structure between triangular to honeycomb, both for the matter-wave and the optical Talbot effect. Sitting on a secondary revival simultaneously for the timing of the matter-wave lens and the positioning of the optical lens could then produce a triangular lattice that would look like a situation for purely primary revivals. We make sure to avoid such a scenario by careful cross-checks.

\section{Experimental setup and calibrations}\label{sec:experiment}
\subsection{Experimental Setup}
The measurements were performed on a lithium quantum gas apparatus with a compact setup. The BEC of $^7$Li atoms in the $|F=1, m_F=1>$ state is prepared in a crossed optical dipole trap (ODT) of three non-interfering laser beams with a wavelength of $\lambda_{\rm lat} = \SI{1064}{nm}$ with angles of about \SIlist{0;120;240}{\degree} in the xy-plane. With a double-pass acoustooptic modulator (AOM) setup we detune the frequency of the single beams by a minimum of $\SI{14}{MHz}$ to each other, which is much larger than all other relevant frequencies in the ODT. The atoms now only see a time averaged potential which has an aspect ratio of $ \omega_z=\sqrt{2}\omega_r$. During evaporative cooling we are working at a magnetic field of \SI{702}{G}. 

For the optical lattice we use the same three beams as for the ODT. To load a triangular lattice we bring all three beams into resonance, i.e., to the same frequency. In addition to the AOMs, we use electrooptic modulators (EOM) to lower the intensity of the carrier, which allows us to control the lattice depth independently of the external confinement. The lattice setup with the double-pass AOMs and EOMs is designed to realize a tunable optical lattice via a multi-frequency lattice \cite{Kosch2022}. The lattice loading sequence is performed in the following way: During evaporation the frequencies  are set to $14$, $0$, $-14$\,MHz with respect to a central frequency. Prior to loading the lattice, we shift them closer to resonance at $+8$, $0$ and $-6$\,MHz avoiding resonance with the closest sidebands of the EOMs at multiples of $10$, $35$ and $45$\,MHz. The power in the carriers is reduced to \SI{10}{\percent} of its unmodulated value in \SI{10}{ms}. Then the frequencies are switched to resonance via the AOMs, thus switching on a very shallow lattice of about $0.14 \, E_{\rm rec}$. In \SI{80}{ms} the overall intensity of the laser beams is increased to about \SI{200}{mW} per beam resulting in a lattice depth of $1.7 \, E_{\rm rec}$. For matter-wave magnification and coherence imaging we need to work close to zero interactions. Therefore we reduce the magnetic field from \SI{702}{G} to \SI{580}{G} and with that the scattering length from $a_s = 76.1 \, a_0$ to $a_s = 3.6 \, a_0$ \cite{Hulet2020}. The lattice depth is then ramped up to its final value ranging from $1.7 \, E_{\rm rec}$ up to $3.7 \, E_{\rm rec}$ in \SI{10}{ms} by decreasing the radio frequency power on the EOMs while the external confinement is only weakly increased due to the stronger interference factor.

We measure a radial confinement from a single lattice beam at this intensity of $\omega_{\rm sb}=2\pi \times \SI{543 (18)}{Hz}$, resulting in a vertical confinement from the three lattice beams of $\omega_z=\sqrt{3}\omega_{\rm sb}=2\pi \times \SI{941(31)}{Hz}$ and an in-plane confinement of $\omega_r=\sqrt{(3/2)}\omega_{\rm sb}=2\pi \times \SI{665(22)}{Hz}$.
The system is held at the final lattice depth for \SI{5}{ms} for thermalization \cite{Schweikhard2007} before imaging it with the matter-wave microscope. The ramp and thermalization times are a compromise with the relatively large three-body loss rate of $^7$Li. We find that for only 1~ms hold time, the system has not thermalized and still contains coherence at the deepest lattices. The thermal dephasing time is given by $\hbar/(k_{\rm B}T)$ \cite{Makhalov2019}, which is between 0.2\,ms and 0.6\,ms for our temperatures. The increasing diameter of the cloud with lattice depth due to the increasing interaction strength in the tighter tubes indicates that the system has time to thermalize.

The BEC has an atom number of typically \num{1e4} or 600 in the central tube. The density distribution is imaged via a matter-wave protocol \cite{Asteria2021} with a magnification of 35.3(5), initialized by switching off the interference between the lattice beams and a fast intensity ramp to an isotropic in-plane trap frequency, which yields an effective trapping frequency of $\omega= 2\pi \times \SI{1.25}{kHz}$ for the $T/4$ pulse (trapping period $T = 2\pi/\omega$). Note that an imaging condition can be found for time-dependent trapping frequencies \cite{Asteria2021}. The $T/4$ pulse is followed by a free expansion of \SI{5}{ms} in the weak in-plane confinement of \SI{20}{Hz} from the residual curvature of the magnetic field. It was shown in ref.~\cite{Asteria2021} that an imaging condition can always be found by combining a near quarter-period pulse with a free expansion of finite length and our expansion in the weak confinement therefore does not correspond to a quarter-period pulse. The duration was chosen instead to optimize the signal. The same magnetic field produces a vertical anti-confinement of \SI{28}{Hz}. We then perform absorption imaging for $\SI{4}{\mu s}$ to obtain the magnified image of the density distribution. 

The experiments require an extremely careful characterization of the isotropy of the optical confinement for the lensing pulse. The half width of the Talbot revivals is just $\SI{3}{\mu s}$, i.e., short compared to the quarter period time of $\SI{200}{\mu s}$, such that a percent-level error in the trap frequency can wash out the image or result in different Talbot revivals in the different directions. To calibrate the trap frequency contribution of the three lattice beams, we prepare the BEC in 1D lattices formed by tuning only two of the three beams to resonance and measure Talbot carpets for varying intensities of the third beam during the lensing pulse. This allows us to iteratively find the condition, where all three directions contribute the same confinement. Isotropic confinements are easier to realize in magnetic traps, but they are not compatible with the Feshbach resonance used in this work.

\subsection{Estimate of the matter-wave magnification.} 
We estimate the matter-wave magnification via the observed lattice constant in the images of $\tilde{a}_{\rm lat}=11.2(2)$~pixels (compare Fig.~1) and the camera pixel size $d=\SI{6.5}{\mu m}$. With the optical magnification of $M_{\rm opt}=3$ obtained from the optical setup and a calibration of Bragg peaks, we obtain $M_{\rm mw}=(d \tilde{a}_{\rm lat})/(a_{\rm lat} M_{\rm opt})=35.3(5)$. 

We compare this with the parameters of the matter-wave protocol. The best estimate for the trapping frequency of the lensing pulse comes from the imaging condition at $\SI{200}{\mu s}$ (at deep lattices, where the interaction shift is smallest) resulting in $\omega=2\pi \times \SI{1.25}{kHz}$. For a lensing pulse followed by a free expansion time $t_{\rm ToF}=\SI{5}{ms}$, the absolute value of the matter-wave magnification is given by $M_{\rm mw}\approx \omega t_{\rm ToF}=39$ \cite{Asteria2021}. For a residual confinement of $\omega_{\rm ToF}=2\pi \times \SI{20}{Hz}$ during the only-approximately free expansion time, the matter-wave magnification is given instead by $M_{\rm mw}=(\omega/\omega_{\rm ToF})\sin(\omega_{\rm ToF} t_{\rm ToF})=36.7$. If we take this correction into account, the two methods are consistent within 4\%. 

\subsection{Discussion of the Talbot effect}

The Talbot revivals in Fig.~3 shift to longer evolution times for shallow lattices, which we attribute to the interactions. In a mean-field description, the repulsive interactions effectively reduce the trapping potential of the lensing pulse requiring a longer evolution time \cite{Asteria2021}. The interaction effects are enhanced at shallow lattices both due to the enhanced local densities from the full interference and the more localized envelope of the initial density (Fig.~3C). We fit the positions of the peaks in the lattice contrast and focus the analysis on the data along these lines for the original matter-wave image and the primary Talbot revivals.

Note that we scan the evolution time in the harmonic trap before the matter-wave protocol and therefore obtain the Talbot times corresponding to the original lattice constant. If we scanned the free evolution time after the lensing pulse instead, we would measure the Talbot revivals of the magnified lattice with impractically long Talbot times of 20\,ms. An alternative protocol would be to switch off the harmonic trap for the short additional evolution time before switching it back on for the lensing pulse. This would probe the Talbot effect in free space instead of a harmonic trap. However, we avoided this additional complication, which would have required extremely precise timing of the switching.

The exact shape of the Talbot carpet depends on the localization on the lattice sites in the original image analogously to the slit widths in the optical Talbot effect. Fractional revivals with smaller periods $a_{\rm lat}/n$, where $n \in \mathbb{N}$, are produced at intermediate evolution times, but their appearance requires an initial distribution with feature sizes on the order of $a_{\rm lat}/n$. Such fractional revivals have been observed with a beam of thermal atoms at micro-fabricated material gratings \cite{Nowak1997}, but they are not expected for atoms released from a shallow optical lattice.

\subsection{Talbot effect in a harmonic trap}
While the Talbot effect is usually considered in free-space evolution, the same revivals also appear for an evolution in a harmonic trap, because they rely on the quadratic phase evolution in momentum space, and the harmonic trap only causes a rescaling of the revival times and lattice constants \cite{Asteria2022}. A parabolic potential was also used in \cite{Mark2011}, but for a different Talbot effect inside an optical lattice, where the quadratic phase evolution on the lattice sites gives rise to revivals in momentum space.

The Talbot effect in a harmonic trap and the rescaling of the Talbot time and the lattice constant are discussed in \cite{Asteria2022}. The evolution time in the harmonic trap for reaching the $n$th Talbot revival is $t_{\rm ho}=(1/\omega) \arctan(n\omega T_{\rm Talbot})$ and the lattice constant is scaled by $\cos(\omega t_{\rm ho})$.
This rescaling of the Talbot time is at the level of only $3\%$ for our parameters even for the third primary revival and $5\%$ for the lattice constant. The expected difference between the Talbot times in the three lattice directions from the misalignment of the lattice beams is on the same order of $3\%$.

\subsection{Calibration of the optical lattice.}
The optical lattice potential can be written as 
\begin{equation}
V_{\rm lat}(\boldsymbol{r})=-2 V_0 \sum_{i=1}^3 \cos(\boldsymbol{b}_i \boldsymbol{r})
\end{equation}
where $\boldsymbol{b}_i$ are the reciprocal lattice vectors, defined as $\boldsymbol{b}_i = \boldsymbol{k}_i - \boldsymbol{k}_{i+1}$ with $\boldsymbol{k}_4 = \boldsymbol{k}_1$. Their length is $|\boldsymbol{b}_i|= 2\pi/a_{\rm 1D} = \sqrt{3}(2\pi/\lambda_{\rm lat})$. Throughout the manuscript, we state the lattice depth in units of the recoil energy $E_{\rm rec} = h^2/(2m\lambda_{\rm lat}^2) = h \times \SI{25}{kHz}$, where $h$ is Planck's constant.

The exact timing is very important and we introduce delay lines in the RF signal lines to the EOMs to compensate small delays between the three lattice beams of 220 and 470\,ns.

We calibrate the optical lattice depth using Kapitza-Dirac scattering. We calculate the single-particle tunneling rate $K$ via a band structure calculation and fit it with the formula \cite{Zwerger2003}
\begin{equation}
 K=g E_{\rm rec}\left(\frac{f V_0}{E_{\rm rec}}\right)^{3/4}\exp \left[-2\left(\frac{f V_0}{E_{\rm rec}}\right)^{1/2}\right]   
\end{equation}
with $f=14.03$ and $g=1.58$.
 
\subsection{Parameters of the XY model}
Our system is well described by a classical XY model. The critical temperature $T_{\rm BEC}$ for condensation in a single well is given by $T_{\rm BEC} \approx 0.94 N_0^{1/3} \hbar \bar{\omega}/k_{\rm B}$ \cite{Ketterle1999}, where $N_0\simeq 100\text{\,-\,}600$ is the atom number per tube. The on-site trapping frequency $\hbar \omega_t=3\sqrt{2V_0 E_{\rm rec}}$ is in the range of $2\pi \times 200\text{\,-\,} 140$\,kHz ($1.7$ to $3.7 \, E_{\rm rec}$) and the vertical confinement of the tubes is given by $\omega_z=2\pi \times \SI{941(31)}{Hz}$. Using the geometric mean of the frequencies $\bar{\omega}$ we estimate $T_{\rm BEC} \approx 7 \text{\,-\,} 10\,\mu$K. Therefore we have $T<T_{\rm BEC}$ and the atoms in the wells of the 2D optical lattice may be described by a macroscopic wavefunction. The observation of speckle patterns for deeper lattices supports this phase coherence within the tubes. 

\begin{figure*}[ht]
    \centering
	\includegraphics[width=0.9\linewidth]{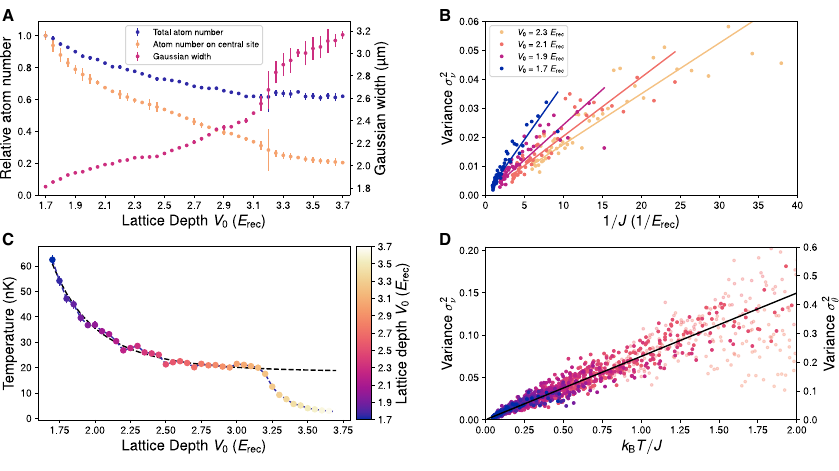}
	\caption{{\bf Calibration of the conversion factor via the temperature dependence of the phase fluctuations.} (A) Relative total atom number $N$ (blue), relative atom number in the central Wigner-Seitz cell $N_0$ (orange) and $1/e^2$-Gaussian width of the cloud (purple) as a function of the lattice depth $V_0$. The maximum values of the atom numbers are $N_{\rm max}=14,743$ and $N_{0,\rm max} = 567$. (B) Measured variance of the relative density fluctuations $\sigma_{\nu}^2$ as a function of the inverse of local tunnel coupling $1/J$ for example lattice depths of $V_0= 1.7, 1.9, 2.1$ and $ 2.3 \, E_{\rm rec}$. We fit the slopes for each lattice depth and extract the temperature, plotted in (C) as described in the main text.  (D) With the temperature curve from (C) the variance data of the relative densities collapse to a single line, when plotted against the scaled temperature $k_{\rm B}T/J$. We fit this slope and enforce $\sigma_{\theta}^2 = 0.22(4) k_{\rm B}T/J$ onto it. With this we get the conversion factor of $\alpha=0.58(6)$.}
   \label{fig:calibration-alpha}
\end{figure*}

The relation $J=2 K N_0$ for the Josephson coupling energy is valid for $J/N_0^2\ll U \ll J$, where $U$ is the Hubbard interaction energy \cite{Trombettoni2005}. With the estimate $U=\frac{4\pi\hbar^2}{m}a_s(\frac{1}{2\pi})^{(3/2)}\frac{1}{a_t^2 a_z}$ with the transverse and vertical harmonic oscillator lengths $a_t$ and $a_z$ we find $U \approx h \times \SI{20}{Hz}$ and the above relations are fulfilled. Furthermore, the interaction is sufficiently small that a shift of the transition temperature is not expected \cite{Smerzi2004}.
The energy of the axial confinement $\hbar \omega_z=h\times \SI{941(31)}{Hz}=k_{\rm B} \times \SI{45}{nK}$ is on the order of the temperatures of $61(2)$ to $19.0(5)$\,nK (see Fig.~\ref{fig:calibration-alpha}). In our range of lattice depths between $1.7$ and $3.7 \, E_{\rm rec}$, the band gap between the first two bands is between 4,\,600 and 43,\,000 times larger than the single particle tunneling energy and thermal excitations into the second band can be completely neglected.

\subsection{Phase thermometry and calibration of the phase microscope}
\label{apx:calibration}
We employ phase thermometry, as previously introduced for a BEC in a double well potential \cite{Gati2006}, to determine both the temperature curve as a function of the lattice depth $V_0$ as well as the conversion factor $\alpha$, by attributing the temperature to the measured thermal phase fluctuations via the relation $\sigma_{\theta}^2(x) \sim k_{\rm B} T/J$, where $\sigma_{\theta}$ is the standard deviation of the phase. We note that such a relation for the phase fluctuations holds generally, but the prefactor depends on the specific system. In 2D, the variance of the phase grows with the logarithm of the system size and it depends on the lattice geometry via the coordination number and dispersion relation \cite{Tobochnik1979}. We have therefore numerically calculated the spin-wave normal modes of the XY model on the real-space triangular lattice of radius 4 sites including the Gaussian envelope of the coupling strength resulting from the density envelope with $1/e^2$ radius of 2.9(4) sites. We then decomposed the fluctuations of single sites into these independently fluctuation normal modes to obtain the relation for the thermal fluctuations $\sigma_{\theta}^2(x)= 0.22(4) k_{\rm B} T/J(x)$ with a constant prefactor, when the phase fluctuations are related to the local tunnel coupling $J(x)$.

We start by computing the relative on-site density fluctuations $\nu_i=(N_i - \bar{N}_i)/\bar{N}_i$ for all lattice sites $i$ where we determine the mean onsite density $\bar{N}_i$ by averaging over 100 images for a given experimental parameter set. We determine an apparent atom-number background standard deviation of 14 atoms per tube from the image regions without atoms, which fits well with an estimate of the photon shot noise of the absorption imaging, and we subtract this value from all fluctuations. We also use this value to calculate the statistical error on the phases given in the main text as the atom shot noise divided by the mean atom number (typically 300 in the center and 70 at the edge) divided by the conversion factor.
The total atom number $N$, the mean atom number in the central Wigner-Seitz cell $N_0$ and the Gaussian width of the cloud are plotted in Fig.~\ref{fig:calibration-alpha}A as a function of the lattice depth $V_0$. 
We then evaluate the variance of those density fluctuations $\sigma_{\nu}^2$ as a function of $1/J=1/(2K\bar{N}_i)$ using the appropriate local atom number $\bar{N}_i$ per tube. Due to the phase microscope, the measured density fluctuations are directly proportional to the thermal phase fluctuations $\sigma_{\nu}^2/\alpha^2=\sigma_{\theta}^2$ with the conversion factor $\alpha$.

For sufficiently small fluctuations i.e. shallow lattices, the variances scale linearly with $T/J$ and thus the slopes $\tau$ of the variances as a function of $1/J$ are proportional to the temperature: $T(V_0)=c_{\tau} \times \tau(V_0)$, with the scaling factor $c_{\tau}$. Examples of this linear scaling together with a linear fit to determine the slope $\tau$ are displayed in Fig.~\ref{fig:calibration-alpha}B.
We fit these slopes $\tau(V_0)$ for each lattice depth and parameterize them as a polynomial in the single particle tunneling rate $K(V_0)$ to extrapolate to larger lattice depths:
$\tau(V_0) = a_0 + a_1 K(V_0) + a_2 K(V_0)^2$ with fit parameters $a_0 = 0.0011 \, E_{\rm rec}$, $a_1 = 1.7$ and $a_2 = 970 \, 1/E_{\rm rec}$. 
This choice is motivated by the expectation of $T(V_0)\propto K$ for a perfectly adiabatic lattice ramp \cite{Blakie2004} and an additional term to account for heating from three-body losses.
Due to the limitations of the phase microscope for large fluctuations, we limit the calibration of the temperature curve to $V_0 \leq 3 \, E_{\rm rec}$.
We use this parameterization to extrapolate to deeper lattices and find the scaling factor $c_{\tau}$ via the condition $(k_{\rm B}T/J)|_{V_0=V_c} = 1.43$. It follows that $c_{\tau} = J/\tau |_{V_0=V_c}\times 1.43 \approx 13.5$, where the largest $J$ in the center of the trap is used at $V_c = 3.25 \, E_{\rm rec}$.  
Fig.~\ref{fig:calibration-alpha}C shows the rescaled slope data as well as the rescaled temperature parameterization. This collapses the variance data to a single line in Fig.~\ref{fig:calibration-alpha}D such that $\sigma_{\theta}^2 = 0.22 \times 1.43=0.31$ at the phase transition. This confirms the expectation of $\sigma_{\nu}^2/\alpha^2=\sigma_{\theta}^2\sim k_{\rm B}T/J$ for thermal phase fluctuations in the XY model \cite{Tobochnik1979}. The condition of $k_{\rm B}T/J=1.43$ at the phase transition at $V_{c}=3.25 \, E_{\rm rec}$ \cite{Okabe2025} which we identified above, fixes both the temperature curve and the conversion factor and we obtain $\alpha=0.58(6)$.

We estimate a systematic relative error on the extracted phases related to the assumption of a constant conversion factor on the order of 9\%, deduced from the standard deviation of the model conversion factors in Fig.~2D in the relevant range of $\beta = 0.33$ and $\beta = 0.5$ corresponding to the distances over which we evaluate the phase coherence function. We note that in future realizations of a phase microscope the calibration of the conversion factor can be achieved without reference to a situation where the physics is known by switching to direct phase imprinting in Fourier space combined with an evolution in a purely harmonic trap \cite{Murthy2019} or by using the Talbot microscope outlined in the outlook. The phase microscope could also be calibrated by comparison to controlled local phase imprinting on individual tubes using optical tweezers, when high optical access is \mbox{available}.

We use here the critical lattice depth $V_c=3.25 \, E_{\rm rec}$ estimated from the decay of the contrast in the Talbot revivals in Fig.~3. However, we note that the identification of $V_c$ does not change the exponents obtained from our analysis in Fig.~4. Replacing $V_c$ by $a V_c$, replaces the fitted temperatures $T$ by $T/a$, i.e. increasing the slope of the variance data by $a$. The conversion factor is obtained via the relation $\sigma_{\theta}^2=\sigma_{\nu}^2/\alpha^2=0.31$ at $k_{\rm B}T/J=1.43$, and $\alpha^2$ is increased by $a$ to keep this relation. The exponent $\eta$ is determined as the slope of $(\Delta\theta)^2$ versus $\log(d)$, which is also scaled by $1/a$. In the final plot of $\eta$ versus $k_{\rm B}T/J$, both axes are scaled by $1/a$ and the slope (which determines the critical exponent $\eta_c$) does not change.

\section{Theory of the phase microscope}
\subsection{Calculation of the matter-wave aberrations}
To estimate the momentum-dependent phase shift from the aberrations we follow the conventions from \cite{Leonard2012}. We expand the trap as $V(x)=(1/2) m \omega_0 x^2 + m(\lambda_4/4) x^4$ and obtain the parameter $\lambda_4=-2(\omega_0/w_0)^2$ for a Gaussian beam profile with $1/e^2$ waist radius $w_0=\SI{41}{\mu m}$. We consider a wave packet initially located in the center of the trap with a momentum $p_\beta=\beta \hbar b$ with the reciprocal lattice vector $b$ of length $b = \sqrt{3}(2\pi/\lambda_{\rm lat})$ and a dimensionless factor $\beta$ that we consider between 0 and 2. This corresponds to an oscillation amplitude $A_\beta=\beta \hbar b/(m \omega_0)$ (with $A_1=\SI{11.8}{\mu m}$) and we introduce the dimensionless amplitude $\alpha_\beta=A_\beta /w_0$ by relating it to the waist $w_0$. We note that $\alpha_\beta<1$ in the relevant range $0<\beta<2$ and we use this to make an expansion in $\alpha_\beta$ below. 

We obtain an amplitude-dependent frequency in the anharmonic trap given by \cite{Leonard2012} $\omega_\beta=\omega_0(1+\epsilon_\beta)$ with $\epsilon_\beta=\frac{3}{8}\frac{A_\beta^2}{\omega_0^2}\lambda_4=-\frac{3}{4}\alpha_\beta^2$ (with $\epsilon_1=-0.062$). The negative quartic part of the potential reduces the trap frequency, or increases the trapping period, such that the quarter-period lensing-pulse time is slightly longer. In the experiment, one would choose the lensing-pulse time to yield the largest lattice contrast. We therefore set the integration time to $\tau=\frac{\pi}{2}\frac{1}{\omega_0(1+\gamma \epsilon_1)}$, i.e. we introduce a dimensionless factor $\gamma$ to parameterize how much the evolution time has to be adapted compared to the case without aberrations. The actual time shift is given by
$\Delta\tau=(\pi/2)(-\epsilon_1/\omega_0)\gamma=\gamma \times \SI{12.4}{\mu s}$
and the value of $\gamma$ is determined below.

The trajectory of the wave packet is given by position $x_\beta(t)=A_\beta \sin(\omega_\beta t)$ and velocity $v_\beta(t)=A_\beta\omega_\beta \cos(\omega_\beta t)$. We neglect the first correction to the position $\frac{\lambda_4 A_\beta^3}{32 \omega_0^2} \sin(3 \omega_\beta t)=- \frac{\alpha_\beta^3 w_0}{16} \sin(3 \omega_\beta t)$, because it leads to terms of order $\alpha_\beta^6$ and higher. The accumulated phase of the wave packet can be calculated from the classical action as $\phi(\tau)=\frac{1}{\hbar}\int_0^\tau (T-V)dt$ with the kinetic energy $T$ and the potential energy $V$. The integral vanishes for quarter-period evolution in a harmonic trap, but we get contributions from the slightly mismatched integration time for $\gamma \neq 0$ and from the quartic potential. We evaluate the integral keeping terms up to order $\alpha_\beta^4$ and obtain 
\begin{widetext}
\begin{align}
\begin{split}
    \phi_\beta&=\frac{1}{\hbar}\int_0^{\tau}[\frac{1}{2}m v^2(t)-\frac{1}{2}m\omega_0^2 x^2(t)+\frac{1}{2}m(\frac{\omega_0}{w_0})^2 x^4(t)]dt\\
    &=\int_0^{\tau}[\frac{m \alpha_\beta^2 w_0^2}{2 \hbar}\omega_\beta^2\cos^2(\omega_\beta t)-\frac{m \alpha_\beta^2 w_0^2}{2\hbar}\omega_0^2\sin^2(\omega_\beta t)
    +\frac{m}{2\hbar}\omega_0 w_0^2 \alpha_\beta^4\sin^4(\omega_\beta t)]dt
\end{split}
\end{align}
Introducing $\tilde{\phi}=\frac{m w_0^2\omega_0}{2\hbar}$ and substituting $u=\omega_\beta t$, $du=\omega_\beta dt=(1-\frac{3}{4}\alpha_\beta^2)\omega_0 dt$ and integrating to $\omega_\beta \tau=\frac{\pi}{2}(1-\frac{3}{4}\alpha_\beta^2+\frac{3}{4}\gamma \alpha_1^2)$
\begin{align}
    \phi_\beta=\tilde{\phi} \alpha_\beta^2 \frac{\omega_\beta}{\omega_0}\int_0^{\omega_\beta \tau} \cos^2(u)du
    - \tilde{\phi} \alpha_\beta^2 \frac{\omega_0}{\omega_\beta}\int_0^{\omega_\beta \tau} \sin^2(u)du +\tilde{\phi} \alpha_\beta^4 \frac{\omega_0}{\omega_\beta}\int_0^{\omega_\beta \tau} \sin^4(u)du.
\end{align}
Expanding the trigonometric functions to quadratic order in $\alpha_\beta$ or $\alpha_1$ around the integration bound, we find
\begin{align}
\begin{split}
    \phi_\beta=\tilde{\phi}\alpha_\beta^2 (1-\frac{3}{4}\alpha_\beta^2)\frac{\pi}{4} - \tilde{\phi}\alpha_\beta^2 (1+\frac{3}{4}\alpha_\beta^2)\frac{\pi}{4}(1-\frac{3}{2}\alpha_\beta^2+\frac{3}{2}\gamma\alpha_1^2) 
    +\tilde{\phi}\alpha_\beta^4[\frac{3\pi}{16}-\frac{\pi}{2}\frac{3}{4}\alpha_\beta^2+\frac{\pi}{2}\frac{3}{4}\gamma\alpha_1^2]\\
    =-\frac{3\pi}{8}\alpha_1^2 \alpha_\beta^2 \tilde{\phi}\gamma+\frac{3\pi}{16}\alpha_\beta^4\tilde{\phi}.
\end{split}
\end{align}
\end{widetext}
The restriction to terms up to $\alpha_\beta^4$ leaves us with a negative term from the mismatched integration time in the harmonic trap and a positive term from the integration over the quartic trap.

We now consider the dependence on momentum via the factor $\beta$. From $\alpha_\beta=\alpha_1\beta=0.283\beta$ and $\tilde{\phi}=742.1$ we obtain 
\begin{equation}\label{eq:phi-supmat}
    \phi_\beta=-\frac{3\pi}{8}\alpha_1^4\tilde{\phi}\gamma\beta^2+\frac{3\pi}{16}\alpha_1^4\tilde{\phi}\beta^4=-1.784\pi\gamma\beta^2+0.892\pi\beta^4
\end{equation}
The lensing-pulse duration parameterized via $\gamma$ is chosen such that the lattice has maximal contrast. This is the case, when the phase difference between momentum 0 and $b$ vanishes, i.e., when $\phi_0=\phi_1$. This is fulfilled for $\gamma=0.5$, i.e. a lensing-pulse duration that is a fraction of $-\gamma\epsilon_1=0.031$ longer than the quarter period. This matches well to the fraction of $0.036$ longer time that we find in the numerical simulations. The phase has a local minimum of $-0.24\pi$ at $\beta=1/\sqrt{2}$. The shape of the accumulated phase versus momentum is shown in Fig.~2C in the main text. 

\subsection{Discussion of the aberrations}
We discuss the dependence of the phases on the system parameters. The shape of the curve is generic for Gaussian beam traps, but the value of the phases scales as $b^4/(m^3\omega_0^3 w_0^2)$. We note that the significant phases that we employ here stem from the choice of the small waist $w_0$ and the light mass $m$ of lithium atoms. There is also a strong dependence on the momenta involved, given by the reciprocal lattice vector $b$, and only small effects are to be expected for large-scale structures such as density waves in dipolar superfluids or few-micron lattice constants, while sub-wavelength lattices would obtain large effects.

We also note that the strength of the aberrations can be tuned for a fixed waist $w_0$ and reciprocal lattice vector $b$ via the choice of the trap frequency $\omega_0$. Counter-intuitively, larger trapping frequencies (with larger resulting magnifications) lead to less aberrations, because the momenta of the initial wave function $p$ are mapped to smaller positions $x = p/(m \omega_0)$ at the end of the lensing pulse, where they experience less aberrations. In our case, the reciprocal lattice vector of length $b$ is mapped to $A_1 = \hbar b/(m \omega_0) = \SI{11.8}{\mu m}$ for our trapping frequency of \SI{1.25}{kHz}, which is small compared to the waist radius of $\SI{41}{\mu m}$ and produces the aberrations shown in Fig.~2. 

We note that matter-wave aberrations can also distort the density distribution even for flat phase profiles. However, for smooth initial density profiles as we use in this article, such effects are negligible and we identify here a parameter regime of aberrations, in which the effect on the phase mapping is effective, while other distortions on the density are negligible.

\subsection{Estimate of the field of view and resolution}
We repeat the above calculation including an initial displacement of amplitude $B$ again neglecting cross terms in the integrals and we find
\begin{equation}
        \phi_\beta=\frac{3\pi}{16}\tilde{\phi}^B\alpha_1^2-(\tilde{\phi}_1+3\tilde{\phi}^B)\frac{3\pi}{16}\alpha_1^2\beta^2+\frac{3\pi}{16}\alpha_1^2\tilde{\phi}_1\beta^4,
\end{equation}
where we have introduced $\tilde{\phi}^B=\frac{m B^2 \omega_0}{2\hbar}$, which is quadratic in the displacement $B$ and therefore always positive. The displacement produces a phase offset without physical consequences and increases the prefactor of the quadratic term. This modification is small as long as the radius of the cloud is $B<A_1/\sqrt{3}=\SI{6.8}{\mu m}$. The field of view of the phase microscope is therefore significantly larger than our system size with typically $B=\SI{2}{\mu m}$ radius. Note that the field of view considered here is not fundamental to the phase microscope if it uses phase imprinting in the Fourier plane and avoids aberrations. 

The matter-wave microscope does not have a clear aperture as optical imaging systems do. Therefore there is no diffraction limit in direct analogy to optics. However, there is a resolution limit given by the matter-wave aberrations. The structure size of the system can be attributed with a momentum and this structure is resolved as long as the absolute value of the aberration phase on this momentum is below about $\pi/4$. For our relatively strong aberrations, the aberration phase stays below this value up to a momentum slightly above the reciprocal lattice vector ($\beta=1.1$), which means that the lattice can be resolved. The apparent size of the Wannier functions is limited by these matter-wave aberrations and not the optical resolution of the absorption imaging.  

\subsection{Discussion of the second image in the incoherent regime}
We discussed above that the direct image of the lattice is defined by a vanishing aberration phase at the reciprocal lattice vector, i.e. $\phi_1=0$, which is reached for a lensing pulse which is slightly longer than the quarter-period pulse by $-\epsilon_1\gamma$ with $\gamma=0.5$. This is the correct argument for a coherent situation, where the wave packets interfere to produce a high density at $\beta=0$ and $\beta=1$ in the Fourier plane. In the case of random phases, the density is instead distributed over many momenta between 0 and 1. One therefore expects a sharp image for an aberration phase profile that is as flat as possible at the cost of a finite value at $\beta=1$. We therefore obtain a second "incoherent" image at around $\gamma=0$ (see Fig.~3A). This is $-0.5\epsilon_1 =\SI{6.2}{\mu s}$ before the first image, which happens to be close to the minus first secondary Talbot revival. It matches with the experimentally observed timing of $\SI{6.5}{\mu s}$ before the first image. When repeating our numerical simulations with random phases, we also find a signal at the reciprocal lattice vector at the respective timing.

The second image is not a Talbot revival, because the observed images at this timing show a triangular lattice and not a honeycomb lattice as would be expected for a secondary Talbot revival. Furthermore, we expect the thermal coherence length to drop below a lattice constant and we also do not see a signal at the timing for the plus first secondary Talbot revival. Instead it is an artifact from the aberrations as explained above.

\subsection{Derivation of the phase microscope in an optical lattice}\label{sec:aberration-phase-microscope}

We now derive the phase microscope for a modulated wave function in an optical lattice with reciprocal lattice vector $b$ ($b=\sqrt{3}(2\pi/\lambda_{\rm lat})$ in the triangular lattice). We start with the case of a constant phase $\phi$ imprinted in momentum space for non-zero momentum. The Bloch wavefunction of quasimomentum zero is given by $\psi(x)=\sum_j c_j e^{ijbx}$, which we restrict to the first two Bloch coefficients $c_0$ and $c_1=c_{-1}$. We can then write the wave function including phase fluctuations $\theta(x)$ as $\psi(x)\propto e^{i\theta(x)}[c_0+c_1 e^{ibx}+c_{-1} e^{-ibx}]=e^{i\theta(x)}[c_0+2c_1 \cos(bx)]$. We apply the phase shift $\phi$ relative to the zero momentum component as above. The density becomes
\begin{align}\label{eqn: phase imprint lattice}
\begin{split}
    n(x) &=|\psi(x)|^2 
       \\&\propto (c_0+2c_1\cos(bx))^2-2\sin(\phi)c_0^2\theta(x),
\end{split}
\end{align}
where we neglected the terms quadratic in $\theta$ as well as a term $\theta(x)\cos(bx)$, which drops out when evaluating the density per Wigner-Seitz cell, and approximating $\cos(\phi)\approx 1$.
Eq.~(\ref{eqn: phase imprint lattice}) yields the conversion factor from initial phase fluctuations to measured relative density fluctuations of $\alpha=-2\sin(\phi)c_0^2$. This factor is reduced by $c_0^2$ compared to a homogeneous BEC, because the lattice distributed the signal over higher momenta that interfere separately. We also note that we needed to assume a small phase shift $\phi \ll \pi/2$ in order to avoid significant density deformations of the lattice. Without the small-angle approximation, the first term in Eq.~(\ref{eqn: phase imprint lattice}) reads $c_0^2+4c_0 c_1 \cos(bx)\cos(\phi)+4c_1^2\cos^2(bx)$ and for $\phi=\pi$, the lattice would appear shifted by half a lattice constant. 

\begin{figure*}[ht]
    \centering
    \includegraphics[width=0.8\linewidth]{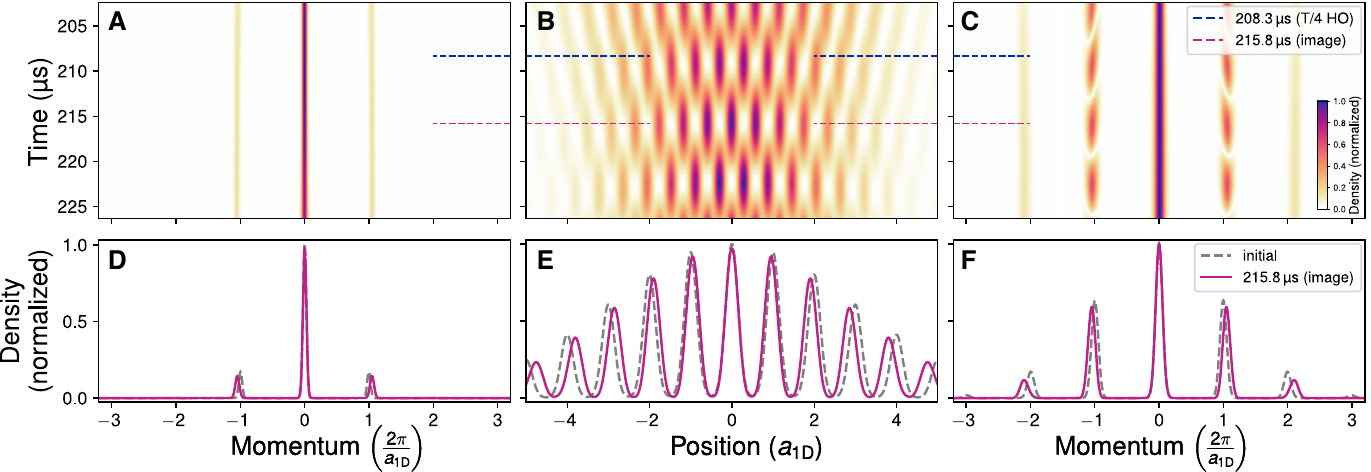}
    \caption{{\bf Numerical simulation of the aberration-induced phase microscope.} 
    (A) Density of the wave function after evolution in the Gaussian trap for times around the $T/4$-time in a harmonic trap ($t_{\rm lens} = \SI{208.3}{\micro s}$, blue dashed line) and the imaging condition including the aberrations ($t_{\rm lens}= \SI{215.8}{\micro s}$, red dashed line). The horizontal axis is expressed in units of momenta to emphasize that Fourier space is realized and for better comparison with (C). (B) Density of the wave function, when the stated time evolution in the Gaussian trap is continued by a time evolution in a harmonic trap for $t_{\rm lens} = \SI{208.3}{\micro s}$ to complete the matter-wave protocol. The imaging condition leads to a sharp lattice structure. The sharp lattice structures shifted out of phase, which occur around $\pm\SI{6.5}{\micro s}$ from the imaging condition, are the first secondary Talbot revival (positive) and its negative counter-part. (C) The absolute value of the Fourier transformation of the final density shows the strongest Bragg peaks for the sharp lattice structures. In between, the Bragg peaks vanish in a way that gives rise to a double peak, which we identify as a finite size effect. The two halves of the peaks are out of phase and can cancel each other out. 
    (D) Density profile at the evolution time of \SI{215.8}{\micro s} in the Gaussian trap (purple) and Fourier transformation of the initial density (gray dashed). (E) Density profile after the full matter-wave protocol (purple) and the initial density for comparison (gray dashed). The Wannier functions on the lattice sites are broadened due to the aberrations in the matter-wave protocol. The matter-wave image is slightly demagnified due to the longer evolution time to compensate for the aberrations. (F) The absolute value of the Fourier transformation of the density profiles in (E) also illustrate the slight demagnification. 
    }
    \label{fig: simulations}
\end{figure*}

We now consider the phase microscope with a momentum-dependent phase shift $\phi_\beta$ at momentum $\beta b$, as imprinted by the matter-wave aberrations discussed above. We write the phase fluctuation profile as a Fourier decomposition 
\begin{equation}
\theta(x)=\int_{\beta=0}^1\left[ a_\beta \cos(\beta bx)+ b_\beta \sin(\beta bx)\right] d\beta
\end{equation}
neglecting a global phase. Assuming small amplitudes $a_\beta$, $b_\beta$ of the phase fluctuations, we can write
\begin{widetext}
\begin{align*}
    \psi(x)&\approx [c_0 + 2c_1\cos(bx)][1 + i \theta(x)]=c_0 + 2c_1\cos(bx) + i c_0\theta(x) + i 2c_1 \theta(x) \cos(bx)\\
    &=c_0 + 2c_1\cos(bx) + i \int_{\beta=0}^1 \left[a_\beta c_0\cos(\beta bx) + b_\beta c_0\sin(\beta bx) \right.
    \\&\phantom{=c_0 + c_1\cos(bx) + i \int_{\beta=0}^1 [}\left.+ a_\beta 2c_1 \cos(bx)\cos(\beta bx) + b_\beta 2c_1\cos(bx)\sin(\beta bx)\right]d\beta\\
    &\rightarrow c_0 e^{i\phi_0} + 2c_1 e^{i\phi_1}\cos(bx)
    \\&\phantom{\rightarrow} + i \int_{\beta=0}^1 \left[c_0 e^{i\phi_\beta} [a_\beta \cos(\beta bx) + b_\beta \sin(\beta bx)] \right.
    \\
    &\phantom{\rightarrow+ i \int_{\beta=0}^1 [}+c_1 e^{i\phi_{1+\beta}}[a_\beta\cos((1+\beta)bx)+b_\beta\sin((1+\beta)bx)] 
    \\&\phantom{\rightarrow+ i \int_{\beta=0}^1 [}\left.+c_1 e^{i\phi_{1-\beta}}[a_\beta\cos((1-\beta)bx)-b_\beta\sin((1-\beta)bx)]\right]d\beta \text{,}
\end{align*}
where we have imprinted the momentum-dependent phases. After setting $\phi_0=0$, neglecting terms quadratic in $a_\beta$ and $b_\beta$ and terms with wave vector outside $[0,b]$, we obtain a density
\begin{align*}
    |\psi(x)|^2&=c_0^2 + 4 c_0 c_1\cos(\phi_1)\cos(bx)+4c_1^2\cos^2(bx)+\int_{\beta=0}^1 \alpha_\beta \left[a_\beta\cos(\beta bx)+b_\beta\sin(\beta bx)\right]d\beta\\
     &=[c_0 + 2c_1\cos(bx)]^2+\int_{\beta=0}^1 \alpha_\beta \left[a_\beta\cos(\beta bx)+b_\beta\sin(\beta bx)\right]d\beta
\end{align*}
with the momentum-dependent conversion factor
\begin{align}
\label{eq:conversion}
    \alpha_\beta=-2c_0^2 \sin(\phi_\beta)-2c_1^2\sin(\phi_{1+\beta}-\phi_1)-2c_1^2\sin(\phi_{1-\beta}-\phi_1).
\end{align}
\end{widetext}
In the last step we have used $\cos(\phi_1)=\cos(0)=1$ as we derived in Eq.~(\ref{eq:phi-supmat}) for our aberrations and indeed this reproduces the original lattice with full contrast. The lattice modulation produces two terms with the momentum-dependent phase fluctuations shifted by plus and minus the reciprocal lattice vector, which obtain the larger aberration-induced phases at these momenta. These terms then interfere with the BEC at plus and minus the reciprocal lattice vector and contribute to the interference signal. For the momentum-independent phase imprinting considered above, these additional terms vanish. For our case, the resulting conversion factor is the sum of these terms and can therefore produce a relatively flat dependence on momentum. Note that the conversion factor $\alpha_\beta$ is positive, because the phases $\phi_\beta$ are mostly negative.

\subsection{Numerical simulations of the phase microscope}
\label{apx: simulation}

We compare the analytical results with numeric simulations by evolving an initial system with the Gross-Pitaevskii equation. The propagation is performed with the split-step method. We use a large real space grid of $10^5$ points with a spacing of \SI{2}{nm} to minimize boundary effects in the momentum domain. For simplicity we work in SI-units and we do not see an accumulation of errors due to small floating point numbers. We simulate the evolution close to our experimental parameters. The trap frequency of the harmonic potential is set to $\omega_0 = 2\pi\times\SI{1.2}{kHz}$. For the anharmonic potential a Gaussian form is used to simulate the light potential. In particular we use $V(x) = V_1\exp(-2x^2/w_0^2)$ with $V_1=-(1/4) m\omega_0^2 w_0^2$ chosen to yield the trapping frequency $\omega_0$ in the harmonic approximation. The waists $1/e^2$-radius of the laser beam is $w_0 = \SI{41}{\micro m}$. We model the initial wavefunction as a sum of equidistant Gaussian wavepackets with a spacing of $a_{\rm 1D} = \SI{614}{nm}$ and a width $\sigma_{\rm wp} = \SI{130}{nm}$, i.e. about the on-site harmonic oscillator length for the relevant lattice depth of $1.7 \, E_{\rm rec}$ to $3.7 \, E_{\rm rec}$. To model the system size of the experiment, we choose a Gaussian density envelope of width $\sigma_{\rm env} = 3 \, a_{\rm 1D}$. The resulting initial density is plotted in Fig.~\ref{fig: simulations}E.

We use a matter-wave protocol consisting of an evolution in the anharmonic trap as described above and a second quarter-period propagation in a purely harmonic trap of the same trapping frequency $\omega_0$ in order to obtain the same aberrations as in the experiment. We replace the time-of-flight expansion of the experiment by the quarter-period propagation in the harmonic trap, because it is numerically less expensive and conceptually equivalent. From the analytic calculation of the aberrations we expect the propagation time in the anharmonic trap for a sharp image to be about by 0.031 longer than the ideal quarter period $t_{\rm lens} = (1/4)(2\pi/\omega_0) = \SI{208.3}{\micro s}$. We therefore repeat the simulation for different propagation times in the anharmonic trap and add the propagation of $t_{\rm lens}$ in the harmonic trap for each case. Fig.~\ref{fig: simulations}A shows the resulting densities at the end of the first and the second propagation together with a Fourier transformation of the latter. We identify the propagation time for the sharp image to be lensing time $t_{\rm lens} = \SI{215.8}{\micro s}$, i.e. by 0.036 longer in agreement with the expectation from the analytic calculation. This yields a demagnification of the lattice of 0.96 that we take into account in the analysis below.

\begin{figure}[hb]
    \centering
    \includegraphics[width=0.9\linewidth]{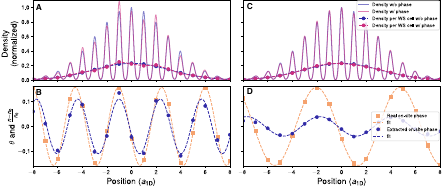}
    \caption{{\bf Numerical determination of the momentum dependent conversion factor.} 
    (A) Determination of the conversion factor for an example phase profile of wavevector $k_\theta = 2\pi/(3.5 a_{\rm 1D})$. Density after the matter-wave protocol with the phase profile (purple) and with constant phase (blue) to calculate the deviation. The data points give the densities integrated over the Wigner-Seitz cells (adapted to the slight demagnification). (B) Phase profile extracted from the relative density profile (purple) together with the initial density profile (blue) (adapted to the slight demagnification). We extract the conversion factor $\alpha=0.7$ to make the two amplitudes match. The slight shift in the wavevector is also visible. (C),(D) same as (A),(B) for a wavevector of $k_\theta = 2\pi/(6.8 a_{\rm 1D})$ yielding a conversion factor of 0.25.
    }
    \label{fig: simulations conversion}
\end{figure}

To simulate the influence of a phase profile, each wavepacket is multiplied with an individual phase and the evolution time in the anharmonic trap set to the imaging condition at $t_{\rm lens} = \SI{215.8}{\micro s}$. The resulting density pattern is compared with that of constant phase. To extract the phase-induced density fluctuations as described in the main text, we integrate over the Wigner-Seitz cells of the final density with the lattice constant adapted to the demagnification, see Fig.~\ref{fig: simulations conversion}. When comparing the extracted phase profile to the original one, we take into account that the matter-wave protocol inverts the image. The extracted density deviations are fitted with a sine. The amplitude of the oscillation compared to the initial phase amplitude gives us the conversion factor. In addition to the slight demagnification, we observe a shift in the wavelength of the phase modulation. For longer wavelengths the extracted modulation wavelength is reduced more strongly. We attribute this to the momentum-dependent demagnification (see momentum-dependent trapping frequency in the calculation of the aberrations).

\subsection{Discussion of the conversion factor}
We can now plug the phase shifts from the matter-wave aberrations of Eq.~(\ref{eq:phi-supmat}) into the expression for the conversion factor for the phase microscope in Eq.~(\ref{eq:conversion}). The resulting shape of the conversion factor versus momentum is shown in Fig.~2D in the main text. It starts at zero, but quickly rises to oscillate around a value of 1 between $\beta=0.3$ and $\beta=0.9$. The oscillation is due to the larger phases in the argument of the second and third term. For the Bloch coefficient $c_0=0.9$ used in the numerics, we obtain a value around 1 in this range.

For the Bloch coefficients for the experimental situation, we perform a band structure calculation and find that the Bloch coefficients are in the intervals $c_0 \in [0.6134, 0.5135]$, $c_1 \in [0.3056, 0.3151]$, and $c_2 \in [0.0885, 0.1376]$ in the range of lattice depths between $1.7$ and $3\, E_{\rm rec}$. An estimate from our derivations of the conversion factor yields $\alpha=-2\sin(\phi_{\rm max})c_0^2=0.43$ for $\phi_{\rm max}=-0.24\pi$ and a typical Bloch coefficient $c_0=0.56$, which fits well to the calibration of $\alpha=0.58(6)$ from the thermal fluctuations and thus corroborates the calculations and the calibration procedure. 

\begingroup
\raggedright
\noindent
Corresponding author: Christof Weitenberg.\\
Email:
\href{mailto:christof.weitenberg@tu-dortmund.de}
{\nolinkurl{christof.weitenberg@tu-dortmund.de}}
\par
\endgroup




%